\documentstyle[aps,  psfig]{revtex}

\begin{document}
\title{Generalized coherent state representation of Bose-Einstein condensates}
\author{V. Chernyak$^{1}$, S. Choi$^{2,3}$, and S. Mukamel$^{2,3}$}
\address{$^{1}$ Corning Incorporated, Process Engineering and Modeling, Corning, New  York 14831 \\
$^{2}$ Department of Chemistry, University of Rochester, Rochester, New York 14627 \\
$^{3}$ Department of Physics and Astronomy, University of Rochester, Rochester, New York 14627}
\maketitle

\begin{abstract}
We show that the quantum  many-body state of Bose-Einstein condensates(BEC) consistent with the time-dependent Hartree-Fock-Bogoliubov (TDHFB) equations is a generalized coherent state (GCS). At zero temperature, the non-condensate density and the anomalous non-condensate correlation are not independent, allowing us to eliminate
one of the three variables in the TDHFB.
\end{abstract}

\pacs{03.75.Fi, 03.75.Dg}

\vspace{6mm}

\section{Introduction}

The recent experimental realization of Bose-Einstein condensates (BEC) in
supercooled trapped atoms has stimulated great interest in the theoretical
description of the quantum state of BEC\cite{history,barnettburnett,lewensteinyou,boldatanwalls,walls,choi}.
The exact quantum state and  many body wavefunction of this system is not known, and simple
approximations such as the
Hartree approximation are commonly employed.  Knowing the quantum state is
crucial for describing the BEC dynamics; in
particular, the many-body hierarchy which leads to an infinite sequence of
progressively higher order equations may be truncated consistently by making an
assumption on the quantum state of the system\cite{mahan}.

Despite the lack of an exact representation of the quantum state,
it has been possible to make significant progress by using physically
sound assumptions about how certain operator products should be factorized
in order to truncate the many-body hierarchy. The Gross-Pitaevskii Equation (GPE)  has been used for describing the dynamics of zero temperature trapped, atomic BEC
with great success\cite{GPE1,GPE2,GPE3,GPE4}. The time-dependent Hartree-Fock-Bogoliubov (TDHFB) equations\cite{HFB1,HFB2,HFB3,HFB4} are
coupled nonlinear equations connecting the dynamics of the condensate
and non-condensate atoms required for the description of finite temperature
BEC in the collisionless regime. Other finite temperature theories include
the time-dependent Bogoliubov-de Gennes equations\cite{CastinDum}, the Quantum Kinetic Theory\cite{Baym,Martin,Gardiner,ZNG,Walser,Zaremba}, and Stochastic methods\cite{Krauth,Ceperley,Walls2,Drummond,Carusotto}.
However none of these treatments directly addresses the precise quantum state of BEC that consists of the condensate as well as the non-condensate atoms. We note that the condensate atoms are described at the mean field level in various theories. However,
 the GPE is derived under the assumption that all correlations of annihilation ($\hat{c}_{i}$) and creation ($\hat{c}_{i}^{\dagger }$) operators for the non-condensate atoms in some basis state $i$ vanish e.g.:
\begin{equation}
\langle \hat{c}_{i}^{\dagger }\hat{c}_{j}^{\dagger }\hat{c}_{k}\hat{c} _{m}
\rangle = \langle \hat{c}_{i}^{\dagger }\hat{c}_{k} \rangle = \langle \hat{c}%
_{j}^{\dagger }\hat{c}_{m} \rangle = 0
\end{equation}
while the HFB equations are derived using a different ansatz:
\begin{equation}
\langle \hat{c}_{i}^{\dagger }\hat{c}_{j}^{\dagger }\hat{c}_{k}\hat{c} _{m}
\rangle =\langle \hat{c}_{i}^{\dagger }\hat{c}_{k} \rangle \langle \hat{c}%
_{j}^{\dagger }\hat{c}_{m} \rangle +\langle \hat{c}_{j}^{\dagger }\hat{c}%
_{k} \rangle \langle \hat{c}_{i}^{\dagger }\hat{c}_{m} \rangle +\langle \hat{%
c} _{i}^{\dagger }\hat{c}_{j}^{\dagger } \rangle \langle \hat{c}_{k} \hat{c}%
_{m} \rangle .  \label{Wick}
\end{equation}

It is straightforward to show that the GPE may be derived by assuming that the quantum state of BEC's at
zero temperature is a coherent state\cite{barnettburnett}. This
is closely related to the description of the laser by a coherent state in Quantum
Optics\cite{glauber1,glauber2}. Indeed one of the earliest stated goals in BEC research
has been the development of an ``atom laser,'' the matter-wave equivalent of
laser\cite{atomlaser1,atomlaser2,atomlaser3}. It has been argued that, owing to the presence of the intrinsic
interatomic collisions, the zero temperature BEC is more accurately
represented by a squeezed state rather than a coherent state\cite{walls,choi}.

Eq. (\ref{Wick}) used for deriving the HFB equations is
reminiscent of Wick's theorem for a system in thermal equilibrium\cite
{mahan,louisell,gaudin,ripka,ring,negele}. This implies that the thermal equilibrium state described by a
statistical density matrix is clearly a possible candidate for the quantum
state of BEC. However, this choice does not
provide a satisfactory physical picture for the TDHFB equations that
describe dynamical condensates away from equilibrium. In addition, collision-induced squeezing\cite{walls} has not been included in the current description of finite temperature BEC. The identification of a quantum state that can describe the dynamics of 
finite temperature BEC, consistent with the TDHFB equations is thus an open issue.

In this paper, we propose a generalized coherent state (GCS) ansatz for the
many body density matrix describing the dynamical quantum state of BEC. Such states were
originally used to describe anharmonic dynamical systems such as many body interacting fermions/bosons\cite{review} while
preserving some of the useful properties of the original Glauber's coherent
states for the harmonic oscillator\cite{glauber1,glauber2}. They encompass the Glauber
coherent state as well as the squeezed state as special cases.
The GCS are particularly convenient for formulating variational dynamics
because of certain algebraic structures originating from the underlying Lie
group algebra\cite{review,perelomov}. Using this ansatz, we  derive the
TDHFB equations\cite{HFB2} via the time-dependent variational principle. This principle which
allows the description of the many-body system in terms of a small number of
parameters is intimately related to classical
Hamiltonian Poisson bracket mechanics that describe
classical dynamics from the minimum action principle.

The paper is organized as follows: In Section II, we review the key
properties of GCS relevant to variational dynamics at zero and finite temperature. In Section III we derive finite-temperature variational equations of motion in both the real space and the trap basis. Conclusions and discussion are given in Section IV.

\section{Generalized coherent state  variational dynamics} \label{Coherent}

Mathematically, a set of GCS is determined by a Lie group $G$, its
irreducible unitary vector representation $T$ with the space $V$ and a
reference state $|\Omega\rangle\in V$. The GCS are defined as states that have a form $T(g)|\Omega\rangle$ with $g\in G$. More specifically, for a generic
quadratic Hamiltonian in some operators $\hat{T}_{i}$
\begin{equation}
\hat{H}=\sum_{i}c_{i}\hat{T}_{i}+\sum_{i,j}c_{ij}\hat{T}_{i}\hat{T}_{j},
\end{equation}
the Lie group $G$ is characterized by the commutation relations amongst the
complete set of operators $\hat{T}_{i}$:
\begin{equation}
[\hat{T}_{i},\hat{T}_{j}]=\sum_{k}C_{ij}^{k}\hat{T}_{k}, \label{commutator}
\end{equation}
where $C_{ij}^{k}$ are known as the structure constants of the set $%
\{\hat{T}_{i}\} $.

For a harmonic oscillator, $\{\hat{T}_{i}\}=\{\hat{a}_{i}, \hat{a}_{j}^{\dagger},\hat{I}\}$ where 
$\hat{a}_{i}$, $\hat{a}_{j}^{\dagger }$ are the boson annihilation and creation operators
and $\hat{I}$ is the identity operator, and $C_{ij}^{k}\equiv \delta _{ij}$, giving the
ordinary Heisenberg-Weyl group. On the other hand, the operator set $\{
\hat{T}_{i} \} = \{ \hat{a}_{i}^{\dagger }\hat{a}_{j},\hat{a}_{i}^{\dagger }\hat{a}_{j}^{\dagger
},\hat{a}_{i}\hat{a}_{j}, \hat{I} \}$  may be used to construct the following Hamiltonian that describes the system
of many-body interacting bosons:
\begin{equation}
\hat{H}_=\sum_{ij}\hat{H}_{ij}\hat{a}_{i}^{\dagger }\hat{a}_{j} + \sum_{ijkl}V_{ijkl}\hat{a}_{i}^{\dagger
}\hat{a}_{j}^{\dagger }\hat{a}_{k}\hat{a}_{l}.  \label{hamiltonian}
\end{equation}
An extended Heisenberg-Weyl algebra may
be obtained by a repeated application of the standard boson commutators $[\hat{a}_i, \hat{a}^{\dagger}_{j}] = \delta_{ij}$.
Writing $\hat{T}^{(-)}_{ij} \equiv \hat{a}_{i}\hat{a}_{j}$, $\hat{T}^{(+)}_{ij} \equiv
\hat{a}_{i}^{\dagger }\hat{a}_{j}^{\dagger }$, $\hat{T}^{(z)}_{ij} \equiv \hat{a}_{i}^{\dagger
}\hat{a}_{j} + \frac{1}{2} \delta_{ij}\hat{I}$, the non-vanishing commutation relations
that define the extended Heisenberg-Weyl algebra are\cite{perelomov}:
\begin{eqnarray}
{\mbox [\hat{T}^{(-)}_{mn}, \hat{T}^{(+)}_{rs}] } & = & \delta_{nr}\hat{T}^{(z)}_{sm} +
\delta_{ns}\hat{T}^{(z)}_{rm} + \delta_{ms}\hat{T}^{(z)}_{rn} + \delta_{mr}\hat{T}^{(z)}_{sn}
\\
{\mbox [\hat{T}^{(z)}_{mn}, \hat{T}^{(-)}_{rs}] } & = & - \delta_{mr}\hat{T}^{(-)}_{ns} -
\delta_{ms} \hat{T}^{(-)}_{nr} \\
{\mbox [\hat{T}^{(z)}_{mn}, \hat{T}^{(+)}_{rs}] } & = & \delta_{nr}\hat{T}^{(+)}_{ms} +
\delta_{ns}\hat{T}^{(+)}_{ms} ,
\end{eqnarray}
while the mixed commutators between the linear and bilinear operators have
the form:
\begin{eqnarray}
{\mbox [\hat{T}^{(+)}_{mn}, \hat{a}_{j}] } & = & - \delta_{mj} \hat{a}_{n}^{\dagger} -
\delta_{nj} \hat{a}_{m}^{\dagger} \\
{\mbox [\hat{T}^{(z)}_{mn}, \hat{a}_{j}] }& = & - \delta_{mj}\hat{a}_{n} \\
{\mbox [\hat{T}^{(-)}_{mn}, \hat{a}_{j}^{\dagger}] } & = & \delta_{nj} \hat{a}_{m} +
\delta_{mj}\hat{a}_{n} \\
{\mbox [\hat{T}^{(z)}_{mn}, \hat{a}_{j}^{\dagger}] } & = & \delta_{nj} \hat{a}_{m}^{\dagger}.
\end{eqnarray}

At zero temperature, the unnormalized generalized coherent states that belong to such extended Heisenberg-Weyl
algebra for interacting bosons have the form\cite{review}:
\begin{equation}
|\psi (\tau )\rangle =\exp \left( \sum_{i}\alpha _{i}(\tau )\hat{a}_{i}^{\dagger
}+\sum_{i,j}\beta _{ij}(\tau )\hat{a}_{i}^{\dagger }\hat{a}_{j}^{\dagger }\right)
|\Omega _{0}\rangle . \label{ansatz}
\end{equation}
The reference state $| \Omega _{0}\rangle$,  which can be normalized to unity $\langle \Omega _{0}|\Omega _{0}\rangle =1$, may be
chosen arbitrarily. However, construction of a useful set of coherent states
for a given dynamic system depends crucially on the choice of $%
|\Omega_{0}\rangle $ which also determines the structure of the phase space
of the dynamical system\cite{review}. We shall take Eq. (\ref{ansatz}) to be our generalized coherent
state ansatz for BEC where the reference state $%
|\Omega _{0}\rangle $ is the particle vacuum state of the ordinary
Heisenberg-Weyl group i.e. $\hat{a}_{i}|\Omega _{0}\rangle =0$. The goal of the variational approach is to determine the time-dependent parameters $\alpha _{i}(\tau)$ and $\beta _{ij}(\tau)$ that represent the evolution of the state $|\psi (\tau )\rangle$
.
Eq. (\ref{ansatz}) indicates that the GCS has a form which combines a
coherent state and a squeezed state.  A coherent state is given by applying a displacement operator $\hat{D}(\alpha)$ on the vacuum $|\alpha \rangle  = \hat{D}(\alpha) |0 \rangle$ where $\hat{D}(\alpha) \equiv \exp \left ( \sum_{i} \alpha^{*}_{i}\hat{a}_
{i} - \alpha_{i} \hat{a}_{i} \right )$. Since $\exp \left (-\alpha^{*}_{i} \hat{a}_{i} \right ) |0 \rangle = |0 \rangle$, an unnormalized coherent state may be written as  $|\alpha \rangle =\exp \left( \sum_{i}\alpha_{i}\hat{a}_{i}^{\dagger} \right ) | 0 
\rangle$. On the other hand, a unitary squeezing operator is given by
$S(\xi)= \exp \left (\frac{1}{2} \sum_{ij} \xi_{ij}^{*} \hat{a}_{i}\hat{a}_{j} - \xi_{ij} a^{\dagger}_{i} a^{\dagger}_{j} \right )$, where $\xi_{ij} = r_{ij} \exp \left ( i \theta_{ij} \right )$ is an arbitrary complex number.

GCS zero temperature variational dynamics is obtained by
implementing the dynamical variational principle assuming that the space of
trial wavefunctions $M$ is represented by a set of GCS. One possible way of
formulating the variational dynamics in Hilbert space is based on projecting
the vector $Hx$ for any $x\in M$ ($H$ being the Hamiltonian operator) into
the tangent subspace to $M$ at $x$. This leads to a vector field in $M$ that
determines the variational dynamics.

The variational equations at zero temperature are derived as follows:
Given a Hamiltonian $\hat{H}$,
and time-dependent wave functions $|\Omega (\tau ) \rangle$, we minimize the action:
\begin{equation}
S[\Omega (\tau )]=\int d\tau \left[ i\left\langle \Omega (\tau )|d\Omega
(\tau )/d\tau \right\rangle -\langle \Omega (\tau )|\hat{H}|\Omega (\tau
)\rangle \right] .  \label{action}
\end{equation}
By choosing a GCS form for $|\Omega (\tau ) \rangle$,
the resulting variational equations can be written in the
Hamiltonian form for any set $\Omega_{j}$ of coordinates which parametrize $%
|\Omega \rangle$:
\begin{equation}
\frac{d \Omega_{j}}{d \tau} = \{ {\cal H}, \Omega_{j} \}
\end{equation}
where $\{ \cdots \}$ denote Poisson brackets and ${\cal H}$ is the classical
Hamiltonian defined by:
\begin{equation}
{\cal H}(\Omega) = \langle \Omega | \hat{H} | \Omega \rangle.
\end{equation}
The use of Poisson brackets clearly establishes the link between the variational equations and the classical dynamics. When the classical Hamiltonian is given by
\begin{equation}
{\cal H} = \sum_{n = 1}^{k} \sum_{i_{1} \cdots i_{n}} h^{(n)}_{i_{1} \cdots
i_{n}} \langle \hat{T}_{i_{1}} \rangle \cdots \langle \hat{T}_{i_{n}} \rangle,
\end{equation}
the Poisson bracket assumes a very simple form provided the wave functions $| \Omega
\rangle$ are parametrized by the expectation values $\langle \Omega | \hat{T}%
_{j} | \Omega \rangle$ of the operators $\hat{T}_{j}$ rather than by the
parameters $\Omega_j$. These  expectation values then constitute a full set of
parameters that uniquely specify the quantum state $|\Omega \rangle $. In
particular, if the operators $\hat{T}_{j}$ form a closed algebra Eq. (\ref{commutator}), the Poisson brackets for $\hat{T}_{j}$ is given by:
\begin{equation}
\{ \hat{T}_{m}, \hat{T}_{n} \} = i \sum_{k} C_{m,n}^{k} \hat{T}_{k},  \label{commutators}
\end{equation}
and the variational equations of motion for $\hat{T}_{m}$ take the closed form
\begin{equation}
i\frac{d \langle \hat{T}_{m} \rangle }{d \tau} = \sum_{n = 1}^{k} \sum_{j=1}^{n}
\sum_{i_{0} \cdots i_{n}} C^{k}_{mi_{0}, \cdots, i_{j}} h^{(n)}_{i_{1}
\cdots i_{n}} \langle \hat{T}_{i_{1}} \rangle \cdots \langle \hat{T}_{i_{n}} \rangle,
\label{variationaleq}
\end{equation}
For the Hamiltonian Eq. (\ref{hamiltonian}), it therefore suffices to
calculate the equations of motion for the expectation values $\langle
\hat{a}_{i}^{\dagger}\hat{a}_{j} \rangle$ and $\langle \hat{a}_{i}\hat{a}_{j} \rangle$ to uniquely specify the
dynamics. In the derivation of the equations of motion we use the
differential property of the Poisson brackets:
\begin{equation}
\{f, gh \} = - \{ gh, f\} = \{ f, g \}h + g \{ f, h \}.  \label{differential}
\end{equation}
When the expectation values are for generators of the set of GCS of some Lie
group $G$, their Poisson brackets are given by the
commutators of the underlying generators of the group. This simplifies the
calculation greatly, as it gives direct correspondence between the ordinary
quantum mechanical commutators and the Poisson brackets. The variational procedure is then formally equivalent to the Heisenberg equations of motion.
It should be noted that the transformation between the expectation values and the parameters may be tedious. However the transformation is never used explicitly; suffice it to know that such transformation exists and we can then proceed to derive closed equations of motion for the expectation values.


The formulation of variational dynamics at finite temperatures constitutes a more
complicated task for the following three reasons: (i) At finite temperatures
the system evolves in {\it Liouville space} and a set of trial density
matrices $M_{L}$ rather than wavefunctions needs to be identified, (ii) an
attempt to project $L\rho$ with $\rho\in M_{L}$ and $L$ is the Liouville
operator into the tangent space to our ansatz faces a difficulty since the Liouville space
does not have a natural scalar product that can be used for this projection
(the scalar product inherited from Hilbert space i.e. the overlap of two
density matrices does not have a direct physical significance), (iii) The
compatibility of the equilibrium and the dynamical approaches that is
straightforward at zero temperature (i.e. the state in $M_{L}$ with the lowest
energy must be the stationary point of the dynamical equation) is not so
obvious in Liouville space. In Appendix \ref{Proof} we show that all of these issues can be adequately resolved for GCS.  In particular we show that, the trial density matrices may be assumed to have the form of finite temperature equilibrium density mar
ices, and that the equilibrium and the dynamical approaches are compatible so that
the trial density matrix which minimizes the Helmholtz free energy is, indeed, a stationary point of the dynamical equations. These results will be used in the coming section.

\section{Variational Equations for Interacting Bosons}

\label{real_space}

\subsection{GCS in real space}

In this section we apply our formalism to a system of
interacting bosons described by the following Hamiltonian with a fixed chemical potential $\mu$:
\begin{eqnarray}  \label{Hamiltonian}
\hat{H} = \hat{H}_{0} + \hat{H}_{f}(t)
\end{eqnarray}
with
\begin{eqnarray}  \label{Ham_int}
\hat{H}_{0}&=& \int d {\bf r} \hat{\psi}^{\dagger}({\bf r}) \left[ -\frac{1}{%
2m}\Delta+V_{trap}({\bf r}) - \mu \right] \hat{\psi}({\bf r})  \nonumber \\
&+&\frac{1}{2}\int d{\bf r}d{\bf r}^{\prime}\hat{\psi}^{\dagger}({\bf r})
\hat{\psi}^{\dagger}({\bf r}^{\prime}) V({\bf r}-{\bf r}^{\prime}) \hat{\psi}%
({\bf r}^{\prime}) \hat{\psi}({\bf r}) \\
H_{f}(t)&=&\int d{\bf r} \hat{\psi}^{\dagger}({\bf r})V_{f}({\bf r}) \hat{%
\psi}({\bf r}).
\end{eqnarray}
where $V_{\rm trap}({\bf r},t)$ is the magnetic potential that confines the atoms and $V_{f}({\bf r},t)$ denotes a general time- and position-dependent external driving potential.
An infinite-dimensional extended Heisenberg-Weyl algebra is generated by the
operators $\hat{\psi}({\bf r})$, $\hat{\psi}^{\dagger}$, $\hat{Y}({\bf r},%
{\bf r}^{\prime}) \equiv \hat{\psi}({\bf r})\hat{\psi}({\bf r}^{\prime})$, $%
\hat{Y}^{\dagger}({\bf r},{\bf r}^{\prime})$, $\hat{N}({\bf r},{\bf r}^{\prime})
\equiv \hat{\psi}^{\dagger}({\bf r})\hat{\psi}({\bf r}^{\prime})$, and $\hat{%
I}$. The space of the representation in which the Hamiltonian [Eq. (\ref
{Ham_int})] is defined can be described as the space of wave-functionals $%
\Psi[x({\bf r})]$, where the one-particle boson operators are given by:

\begin{eqnarray}  \label{boson_oper}
\hat{\psi}({\bf r})=-\frac{i}{\sqrt{2}}\left[ \frac{\delta}{\delta x({\bf r})%
}-x({\bf r}) \right], \;\;\;\;\; \hat{\psi}^{\dagger}({\bf r})=-\frac{i}{%
\sqrt{2}}\left[ \frac{\delta}{\delta x({\bf r})}+x({\bf r}) \right].
\end{eqnarray}
$x({\bf r})$ is a harmonic oscillator coordinate associated with position ${\bf r}$ and Eq. (\ref{boson_oper}) can be used to represent of the operators $\hat{Y}$, $\hat{Y}^{\dagger}$, and $\hat{N}$.

At zero temperature the set of coherent states represented by Gaussian
wavefunctions
\begin{eqnarray}  \label{coherent_func}
\Psi[x({\bf r})]=A\exp\left\{-\frac{1}{2}\int d{\bf r}d{\bf r}%
^{\prime}\sigma({\bf r},{\bf r}^{\prime}) [x({\bf r})-x_{0}({\bf r})][x({\bf %
r}^{\prime})-x_{0}({\bf r}^{\prime})] \right\}
\end{eqnarray}
are parametrized by complex-valued functions $x_{0}({\bf r})$, and $\sigma(%
{\bf r},{\bf r}^{\prime})$.
According to the formalism developed in Section \ref{Coherent} and Appendix \ref{Proof} the coherent finite temperature
density matrices $\rho[x_{j}({\bf r})]$ with $j=L,R$ [Left (ket), Right (bra)] are represented by
Gaussian wavepackets:
\begin{eqnarray}  \label{coherent_dens}
\rho[x_{j}({\bf r})]=Z^{-1}\exp\left\{-\frac{1}{2}\sum_{kj}\int d{\bf r}d%
{\bf r}^{\prime}\sigma_{kj}({\bf r},{\bf r}^{\prime}) [x_{k}({\bf r}%
)-x_{k}^{(0)}({\bf r})][x_{j}({\bf r}^{\prime})-x_{j}^{(0)}({\bf r}%
^{\prime})] \right\} .
\end{eqnarray}
These trial density matrices are parametrized by the
vector and matrix functions $x_{k}^{(0)}({\bf r})$ and $\sigma_{kj}({\bf r},%
{\bf r}^{\prime})$. We note that the Gaussian wavepacket [Eq. (\ref{coherent_dens})] constitutes a coordinate representation for a
density matrix of the form $\rho=Z^{-1}\exp(-K)$, where $K$ is given by a
combination of linear and bilinear terms in single-particle operators.

The variational parameters of Eqs. (\ref{coherent_func}) and (\ref{coherent_dens})  which denote the displacement and the width of the Gaussian wave packet in phase space are related to the average number of particles 
and the quantum mechanical squeezing of the number-phase conjugate variables in BEC.
These parameters may be related to physical quantities such as the condensate fraction and the excitation energy, by transforming Eq. (\ref{coherent_func}) or Eq. (\ref{coherent_dens}) to the quasiparticle basis; the resulting relationship between the variational parameters in different bases is not simple. However this transformation is never used explicitly since the GCS ansatz allows us to derive equations of motion directly for the parameters of interest; the expectation values of the relevant operators.

The most convenient parametrization for trial wavefunctions [Eq. (\ref{coherent_func})] or
density matrices [Eq.  (\ref{coherent_dens})] is given by the expectation values of linear and bilinear
combinations of boson single particle operators:

\begin{eqnarray}  \label{parametrization}
z({\bf r}) \equiv \langle \hat{\psi}({\bf r}) \rangle, \;\;\;\; \kappa({\bf r},{\bf %
r}^{\prime}) \equiv \langle \hat{Y}({\bf r},{\bf r}^{\prime}) \rangle
-z({\bf r})z({\bf r}^{\prime}), \;\;\; \rho({\bf r},{\bf r}%
^{\prime}) \equiv \langle \hat{N}({\bf r},{\bf r}^{\prime})
\rangle -z^{*}({\bf r})z({\bf r}^{\prime})
\end{eqnarray}
where the expectation value is taken with respect to the wavefunctions given by
Eq. (\ref{coherent_func}) or density matrices of Eq. (\ref{coherent_dens}).
Wick's theorem\cite{mahan,louisell,gaudin,ripka,ring,negele} allows us to express the
expectation value of any operator in terms of the parameters given in Eq. (\ref
{parametrization})  both at zero and finite temperature.
The dynamical equations for the system of
interacting bosons may therefore be derived in the same way for both zero and non-zero
temperatures by starting with the  Heisenberg equations of motions for linear and
bilinear combinations of the single-particle operators and then evaluating
the right hand sides using the Wick's theorem. This results in
closed equations for the parameters $z({\bf r})$, $\kappa({\bf r},{\bf r}%
^{\prime})$, and $\rho({\bf r},{\bf r}^{\prime})$, which will be derived next.

\subsection{Variational equations of motion in real space}

The Heisenberg equation of motion for $\hat{%
\psi}({\bf r})$ reads:
\begin{equation}
i \hbar \frac{d\hat{\psi}({\bf r})}{dt}
=  H^{sp} \hat{\psi}({\bf r}) + \int d {\bf r}^{\prime}\hat{\psi}%
^{\dagger}({\bf r}^{\prime}) \bar{V}({\bf r},{\bf r}^{\prime})\hat{\psi}(%
{\bf r}^{\prime}) \hat{\psi}({\bf r})  \label{hatpsidot}
\end{equation}
where
\begin{eqnarray}
H^{sp}({\bf r}) & \equiv & -\frac{1}{2m}\Delta + V_{\rm trap}({\bf r}), \\
\bar{V}({\bf r},{\bf r}^{\prime }) & \equiv & \frac{1}{2}\left[ V({\bf r}-%
{\bf r}^{\prime })+V({\bf r}^{\prime }-{\bf r}) \right ],
\end{eqnarray}
and we have used the commutation relations for the boson field operators
\begin{equation}
[ \hat{\psi}({\bf r}), \hat{\psi}^{\dagger}({\bf r^{\prime}})] = \delta({\bf %
r} - {\bf r}^{\prime}), \;\;\;\; [ \hat{\psi}({\bf r}), \hat{\psi}({\bf %
r^{\prime}})] = [ \hat{\psi}^{\dagger}({\bf r}), \hat{\psi}^{\dagger} ({\bf %
r^{\prime}})] = 0 .
\end{equation}
Taking the expectation values of Eq. (\ref{hatpsidot}) and noting the definition of $\kappa ({\bf r},{\bf r}^{\prime })$ and $\rho ({\bf r},{\bf r}%
^{\prime })$ [Eq. (\ref{parametrization})], we obtain the equation of motion for the mean field:
\begin{eqnarray}
i \hbar \frac{d z({\bf r})}{dt} = H^{sp}({\bf r}) z({\bf r}) & + & \int  d {\bf %
r}^{\prime} \bar{V}({\bf r}, {\bf r}^{\prime}) \left \{ |z({\bf r}%
^{\prime})|^{2}z({\bf r}) + z^{*}({\bf r}^{\prime}) \kappa({\bf r}, {\bf r}%
^{\prime}) \right.  \nonumber \\
& + & \left. z({\bf r}^{\prime}) \rho({\bf r}, {\bf r}^{\prime}) + z({\bf r}%
) \rho({\bf r}^{\prime}, {\bf r}^{\prime}) \right \} + V_{f}({\bf r},t) z({\bf r}) . \label{z_r}
\end{eqnarray}

The equations of motion for $\kappa ({\bf r},{\bf r}^{\prime })$ and $\rho (%
{\bf r},{\bf r}^{\prime })$ can be derived similarly by computing the time derivatives using  Eq. (\ref{parametrization}) and Eq. (\ref{hatpsidot}) in the product rule:
\begin{eqnarray}
i\hbar \frac{d\rho ({\bf r},{\bf r}^{\prime })}{dt} &=&H^{sp}({\bf r})\rho (%
{\bf r},{\bf r}^{\prime })+\int d{\bf r}^{\prime \prime }\bar{V}({\bf r}%
^{\prime },{\bf r}^{\prime \prime })\left\{ \tilde{\xi}({\bf r}^{\prime
\prime },{\bf r}^{\prime })\rho ({\bf r},{\bf r}^{\prime \prime })+ \tilde{%
\xi}({\bf r}^{\prime \prime },{\bf r}^{\prime \prime })\rho ({\bf r},{\bf r}%
^{\prime })\right.  \nonumber \\
&+&\tilde{\zeta} ({\bf r}^{\prime },{\bf r}^{\prime \prime })\kappa ^{\ast }(%
{\bf r}^{\prime \prime },{\bf r})\}-H^{sp}({\bf r}^{\prime })\rho ({\bf r},%
{\bf r}^{\prime })-\int d{\bf r}^{\prime \prime }\bar{V}({\bf r},{\bf r}%
^{\prime \prime })\left\{ \tilde{\xi}({\bf r},{\bf r}^{\prime \prime})\rho (%
{\bf r}^{\prime \prime},{\bf r}^{\prime })\right.  \nonumber \\
&+& \tilde{\xi}({\bf r}^{\prime \prime },{\bf r}^{\prime \prime })\rho ({\bf %
r},{\bf r}^{\prime })+ \tilde{\zeta} ^{\ast }({\bf r},{\bf r}^{\prime \prime
})\kappa ({\bf r}^{\prime \prime },{\bf r}^{\prime })\} + V_{f}({\bf r},t) \rho({\bf r},{\bf r}^{\prime })  \nonumber \\
&-& V_{f}({\bf r}',t) {\rho}({\bf r},{\bf r}^{\prime })  \label{rho_rr}
\end{eqnarray}

\begin{eqnarray}
i\hbar \frac{d\kappa ({\bf r},{\bf r}^{\prime })}{dt} &=&H^{sp}({\bf r}%
)\kappa ({\bf r},{\bf r}^{\prime })+ \int d{\bf r}^{\prime \prime }\bar{V}(%
{\bf r}^{\prime },{\bf r}^{\prime \prime })\left\{ \tilde{\xi}({\bf r}%
^{\prime \prime }, {\bf r}^{\prime })\kappa ({\bf r},{\bf r}^{\prime \prime
})+ \tilde{\xi}({\bf r}^{\prime \prime },{\bf r}^{\prime \prime })\kappa (%
{\bf r},{\bf r}^{\prime })\right.  \nonumber \\
&+&\left. \tilde{\zeta}({\bf r}^{\prime },{\bf r}^{\prime \prime }) \left [
\rho^{*} ({\bf r}, {\bf r}^{\prime \prime }) + \delta ({\bf r} - {\bf r}%
^{\prime \prime }) \right ] \right\} + H^{sp}({\bf r}^{\prime })\kappa ({\bf %
r},{\bf r}^{\prime }) + \int d{\bf r}^{\prime \prime }\bar{V}({\bf r},{\bf r}%
^{\prime \prime })  \nonumber \\
& \times & \left\{ \tilde{\xi}({\bf r}^{\prime \prime },{\bf r})\kappa ({\bf %
r}^{\prime },{\bf r}^{\prime \prime }) + \tilde{\xi}({\bf r}^{\prime \prime
},{\bf r}^{\prime \prime })\kappa ({\bf r},{\bf r}^{\prime }) + \zeta ({\bf r%
},{\bf r}^{\prime \prime })\rho ({\bf r}^{\prime \prime },{\bf r}^{\prime
})\right\} \nonumber \\
&+ & V_{f}({\bf r},t) {\kappa}({\bf r},{\bf r}^{\prime }) + V_{f}({\bf r}',t) {\kappa}({\bf r},{\bf r}^{\prime }) , \label{kappa_rr}
\end{eqnarray}
where we have introduced the auxiliary functions
\begin{eqnarray}
\tilde{\xi}({\bf r},{\bf r}^{\prime }) &=&z^{\ast }({\bf r})z({\bf r}%
^{\prime })+\rho ({\bf r},{\bf r}^{\prime }), \label{xi}\\
\tilde{\zeta} ({\bf r},{\bf r}^{\prime }) &=&z({\bf r})z({\bf r}^{\prime
})+\kappa ({\bf r},{\bf r}^{\prime }) . \label{zeta}
\end{eqnarray}
For the commonly used special case of the contact interatomic interaction for $\bar{V}({\bf r},{\bf r}^{\prime })$, Eqs. (\ref{z_r}-\ref{kappa_rr}) are simplified greatly; these are given in Appendix \ref{simple}.

An important consequence of the GCS ansatz is that at zero temperature the functions $\rho ({\bf r},{\bf r}^{\prime })$ and $\kappa ({\bf r}%
,{\bf r}^{\prime })$ are, in fact, not independent\cite{chernyak}. By deriving an explicit
relationship between them, it is possible to eliminate the $\rho ({\bf r},{\bf r}^{\prime })$ variables. This relation is derived in the
trap basis in Appendix \ref{relrhokappa}, and then converted into the real space basis:


\begin{equation}
\rho ({\bf r},{\bf r}^{\prime })=\left[ \sqrt{\frac{1}{4}\delta (%
{\bf r},{\bf r}^{\prime })+ \int \kappa^{*}({\bf r},{\bf r}'')\kappa ({\bf r}'',
{\bf r}') d {\bf r}'' }-\frac{1}{2}\delta ({\bf r},{\bf r}^{\prime })\right] . \label{rhokappa}
\end{equation}

It can be verified by direct substitution that once Eq. (\ref{rhokappa}) holds initially, it remains true throughout the
dynamical evolution. The reduced set of equations are then the coupled
equations Eqs. (\ref{z_r}) and (\ref{kappa_rr}) with $\rho ({\bf r},{\bf r}%
^{\prime })$ replaced by the expression Eq. (\ref{rhokappa}). The two
independent variables $z({\bf r})$ and $\kappa ({\bf r},{\bf r}^{\prime })$
constitute a very convenient parametrization of squeezed states. $z({\bf %
r})$ represents the average position whereas $\kappa ({\bf r},%
{\bf r}^{\prime })$ are responsible for squeezing. This can be easily understood from the fact that the coherent state  is an eigenstate of the annihilation operator $\hat{\psi}({\bf r})$  while a squeezed state is generated using a squeezing operator which is a function of the quadratic operator $\hat{Y}({\bf r},{\bf r}')$ in the extended Heisenberg-Weyl algebra.

\subsection{Variational equations of motion in the trap basis}

For completeness, we outline below the derivation of the same variational
equations in the trap basis. In this basis, the
Hamiltonian is written as:
\begin{equation}
H=\sum_{ij} H_{ij}\hat{a}_{i}^{\dagger }\hat{a}_{j} + \sum_{ijkl}V_{ijkl}\hat{a}_{i}^{\dagger
}\hat{a}_{j}^{\dagger }\hat{a}_{k}\hat{a}_{l} + \sum_{ij} E_{ij} a^{\dagger}_{i}\hat{a}_{j}. \label{traphamiltonian}
\end{equation}
The matrix elements of the single particle Hamiltonian $H_{ij}$ are given by
\begin{equation}
H_{ij} = \int \! d^{3}{\bf r} \, \phi^{*}_{i}({\bf r}) \left [ - \frac{\hbar^2}{2m}
\Delta + V_{\mbox{\scriptsize trap}}({\bf r}) \right ]  \phi_{j}({\bf r}),
\end{equation}
where the basis state $\phi_{i}({\bf r})$ is arbitrary; a convenient basis for trapped BEC is the eigenstates of the trap since $H_{ij}$ is then diagonal. The indices may also be viewed as the mode indices in a multimode quantum state.
The symmetrized two particle interaction matrix elements are
\begin{equation}
V_{ijkl} = \frac{1}{2} \Big [ \mbox{$\langle i
j |$}V\mbox{$| k l \rangle$} + \mbox{$\langle j i |$}V\mbox{$| k l \rangle$} %
\Big ],
\end{equation}
where
\begin{equation}
\begin{array}{c@{\hspace{1cm}}c}
{\displaystyle \mbox{$\langle i j |$}V\mbox{$| k l \rangle$} = \int \! d^{3}%
{\bf r} \, d^{3}{\bf r^{\prime}} \, \phi^{*}_{i}({\bf r})\phi^{*}_{j}({\bf %
r^{\prime}})V({\bf r}-{\bf r^{\prime}})\phi_{k}({\bf r^{\prime}})\phi_{l}(%
{\bf r}) , } &
\end{array}
\end{equation}
with $V({\bf r}-{\bf r^{\prime}})$ being a general interatomic potential. Also,
\begin{equation}
E_{ij} \equiv \int d {\bf r} \phi^{*}_{i}({\bf r})V_{f}({\bf r},t)\phi_{j}({\bf r}),
\end{equation}
where $V_{f}({\bf r},t)$ denotes a general time- and position-dependent external driving potential as defined previously.
First, we proceed by establishing that our generalized coherent state Eq. (%
\ref{ansatz}) at zero temperature is a Gaussian in coordinate space. With
the choice of the particle vacuum state of ordinary Heisenberg-Weyl algebra as
our reference state $| \Omega_{0} \rangle$, Eq. (\ref{ansatz}) the action of
the operators $\hat{a}_{i}$ and $\hat{a}_{i}^{\dagger }$ on the wave function $%
\Omega (q_{1},\ldots q_{{\cal N}})$ in the coordinate representation, where ${\cal N}$ is
the total number of bosons is:
\begin{equation}
\hat{a}_{i}=-\frac{i}{\sqrt{2}}\left( \frac{\partial }{\partial q_{i}}%
-q_{i}\right) ,\;\;\;\;\;\;\hat{a}_{i}^{\dagger }=-\frac{i}{\sqrt{2}}\left( \frac{%
\partial }{\partial q_{i}}+q_{i}\right) ,  \label{aiq}
\end{equation}
and the conditions $\hat{a}_{i}\Omega (q_{1},\ldots q_{{\cal N}})=0$ for $i=1\ldots {\cal N}$
imply that the reference state $\Omega _{0}(q_{1},\ldots q_{{\cal N}})$ is a
Gaussian in coordinate space:
\begin{equation}
\Omega _{0}(q_{1},\ldots q_{{\cal N}})=\frac{1}{\sqrt{(2\pi )^{{\cal N}}}}\exp \left[ -%
\frac{1}{2}(q_{1}^{2}+\cdots +q_{{\cal N}}^{2})\right] . \label{gauss}
\end{equation}
It should be noted that, consistent with our choice of basis in Eq. (\ref{traphamiltonian}), the 
index $i$ of the coordinate variable $q_i$ in Eqs. (\ref{aiq}) and (\ref{gauss}) refers to the trap 
basis ``mode'' $i$. In addition, although we are using the trap basis, 
the finite total number of particles
${\cal N}$ implies that the vector space used is effectively a finite
dimensional space spanned by a truncated set of trap basis states.   
Acting on this wave function with the generalized displacement operator[Eq. (\ref{ansatz})] preserves its Gaussian form since the action by the
operators $\hat{a}_{i}$ and $\hat{a}_{i}^{\dagger }$ simply shifts the origin while the
operators $\hat{a}_{i}\hat{a}_{j}$ and $\hat{a}_{i}^{\dagger }\hat{a}_{j}^{\dagger }$ change the
variance. The resulting state is thus a Gaussian of the form:
\begin{equation}
\Omega (q_{1},\ldots q_{{\cal N}})=A\exp \left[ -\frac{1}{2}\sum_{i,j}\sigma
_{ij}(q_{i}-\eta_{i})(q_{j}-\eta_{j})\right] .
\end{equation}
where $\eta_{i}$, $i=1\ldots {\cal N}$ are the complex numbers which determine
the average position while $\sigma _{ij}$ is an ${\cal N}\times {\cal N}$ symmetric matrix
that determines the covariances or the amount of squeezing. For an ordinary
coherent state, $\sigma _{ij}=\delta _{ij}$ i.e. a Gaussian
with unit covariance.

Similarly, at finite temperatures, the trial density matrix takes the form ($k,j=L,R$):
\begin{eqnarray}  \label{coherent_dens_trap}
\rho[q^{j}_{1}, \ldots , q^{j}_{{\cal N}}]=Z^{-1}\exp\left\{-\frac{1}{2}\sum_{kj}\sum_{lm} \sigma^{kj}_{lm} [q^{k}_{l}-q^{k(0)}_{l}][q^{j}_{m} -q^{j(0)}_{m}] \right\} .
\end{eqnarray}

The GCS ansatz may therefore be considered to be a ground state of some
effective quadratic Hamiltonian in coordinate and
momentum operators. For such a Hamiltonian, any correlation function can be
represented in a path-integral form where the action only has linear and
bilinear terms\cite{chernyakreferences}. The resulting Wick's theorem is then identical to that for a thermal state.

The expectation values to be used in the parametrization our state are
the condensate mean field $z_{i}$, the non-condensate density $\rho_{ij}$
and the non-condensate correlations $\kappa_{ij}$:
\begin{equation}
z_{i} \equiv \langle \hat{a}_{i} \rangle \;\;\; \rho_{ij} \equiv \langle
\hat{a}_{i}^{\dagger} \hat{a}_{j} \rangle - \langle \hat{a}_{i}^{\dagger}
\rangle \langle \hat{a}_{j} \rangle \;\;\; \kappa_{ij} \equiv \langle \hat{a}%
_{i} \hat{a}_{j} \rangle - \langle \hat{a}_{i} \rangle \langle \hat{a}_{j}
\rangle .
\end{equation}

Since these variables
are the expectation values of the generators of the set of generalized
coherent states of the extended Heisenberg-Weyl algebra, their Poisson
brackets are given by the commutators of the underlying generators.
This results in TDHFB equations of motion
for $z_{i}$, $\rho _{ij}$ and $\kappa _{ij}$.

At zero temperature, the relationship between $\rho_{ij}$ and $\kappa_{ij}$ [Eq. (\ref{rhokappa})] is:
\begin{equation}
\rho_{ij} = \sqrt{\frac{1}{4} \delta_{ij} + \sum_{p} \kappa_{ip}^{*}
\kappa_{pj} } - \frac{1}{2} \delta_{ij} .
\end{equation}
This enables us to reduce the number of equations. More details of this relation are provided in Appendix \ref{relrhokappa}  while the TDHFB equations in the trap basis including the simplified zero temperature form  are given in Appendix \ref{trapTDHFB}.

\section{Discussion}

Using the GCS ansatz, we have derived variationally the TDHFB equations equations of motion for BEC, which are known to be valid in the collisionless regime. This implies that the GCS ansatz should be applicable in the lower temperature, collisionless regime. It should be noted that the HFB theory has several inconsistencies such as the violation of the Hugenholtz-Pines theorem\cite{HP} which states that the excitation spectrum should be gapless in the homogeneous limit\cite{HFB1}. This issue has been addressed by various authors; for instance, the Popov approximation, in which the anomalous correlation is neglected, was shown to give a gapless spectrum\cite{HFB1}. Recently, it has been shown that by replacing the contact interaction potential with a more
 sophisticated pseudopotential, many of the inconsistency problems of the HFB equations including the violation of the Hugenholtz-Pines theorem, inconsistencies with the many body $T$-matrix calculations, and the ultraviolet !
divergences can be overcome\cite{UV}. In this paper we have presented our HFB equations with the general interaction in both the real space and the trap basis; pseudopotentials such as those discussed in Ref. \cite{UV} can thus be accommodated.

Since the GCS is a squeezed state, the present work may be considered an extension of a previous
result that demonstrated {\em stationary} BEC to be squeezed\cite{walls} and
a more recent result that has shown that dynamically evolving BEC under the
time-dependent GPE described using the Hartree approximation (i.e. pure
condensate, no non-condensate atoms) is squeezed\cite{choi}.

The representation of the dynamical quantum state of BEC as a GCS provides physical insight about the {\em total}
system of condensates plus non-condensates in terms of particle annihilation
and creation operators, and how it evolves as a whole in the Schr\"{o}dinger
picture. The interdependence of $\rho $ and $\kappa $ at zero temperature enables us to
eliminate the $\rho$  variables from these equations, reducing the size
of the problem and simplifying the numerical solution.

The multimode squeezing, and hence the entangled state nature of BEC is
clear from the form of the quantum state, Eq. (\ref{ansatz}). The study of
quantum entanglement is currently gaining great interest owing to its
importance in quantum information theory\cite{QI1,QI2} as well as in the understanding of the foundations of quantum mechanics\cite{foundations1,foundations2}. Experimental sources of generalized coherent states in
matter-waves already exist in the form of atomic BEC's. However, ways to access and manipulate this type of matter-wave entanglement remains an open challenge.

\acknowledgments
The support of NSF Grant No. CHE-0132571 is gratefully acknowledged.

\appendix

\section{Properties of finite temperature GCS} \label{Proof}

As noted in the main text, the formulation of variational dynamics at finite temperatures requires us to address the following issues: (i) At finite temperatures
the system evolves in Liouville space and a set of trial density
matrices $M_{L}$ need to be identified, (ii) an
attempt to project $L\rho$ with $\rho\in M_{L}$ and $L$ is the Liouville
operator into the tangent space to our ansatz faces a difficulty since the Liouville space does not have a natural scalar product that can be used for this projection, (iii) The compatibility of the equilibrium and the dynamical approaches is not so obvious in Liouville space. In this Appendix, we show how all these issues may be adequately addressed. 
We start by introducing the trial density matrices.

Let ${\cal A}$ be the real Lie algebra of the real Lie group $G$ involved in the definition
of a set of GCS.  For our case, ${\cal A}$ is the basic (real) algebra generated by the generators $\hat{a}_{j}$, $\hat{a}^{\dagger}_{j}$, and $\hat{T}_{mn}$; the elements $a$ belonging to the algebra ${\mathcal A}$, $a\in{\mathcal A}$, are then linear combinations of these generators. 
In addition, let ${\mathcal A}^{(c)}$ and $G^{(c)}$ be the complexification of ${\cal A}$ and $G$ i.e. $G^{(c)}$ is the complex Lie group that corresponds to the complex Lie algebra ${\mathcal A}^{(c)}$. Complexification of an algebra (group) gives an 
algebra (group) generated by the original generators for which the coefficients are allowed to be complex, rather than real numbers. More specifically, it means constructing a complex analytical
algebra (group), for which the ``real'' version is the original one. For example, given a real structure which is a map $p:G \rightarrow G$ the real part of the group consists of points $g$ so that $p(g)=g$. If one defines $p(g)=g^{\dagger}$ one has $G_{r
eal}=SU(2)$, while if one defines $p(g)=g^*$, one  has $G_{real}=SL(2,R)$. $SL(2,C)$ is then a 3-dimensional complex analytical group (3 is its complex dimension) which serves as complexification for both $SU(2)$ and $SL(2,R)$.

The representation $T$ of a group $G$ associates with any $g \in G$ a linear operator $T(g)$ (acting in some complex vector space referred to as the space of the representation) so that $T(g_{2}g_{1})=T(g_{2})T(g_{1})$.  $T$ can be naturally extended to a
 representation of $%
G^{(c)}$ in the same vector space $V$.  A representation of a group has the corresponding representation of an algebra and vice versa, and in the corresponding algebra representation we have $T(a)$ being operators in the same space for $a \in {\mathcal A}
$ with $T(a_{1}+a_{2})=T(a_{1})+T(a_{2})$ and $T([a_{1},a_{2}])=T(a_{1})T(a_{2})-T(a_{2})T(a_{1})$.  The representation of ${\mathcal A}$ can also be easily extended to representation of ${\mathcal A}^{(c)}$.  

We define the manifold $M_{L}$ of
normalized trial density matrices represented by $\rho(g)=Z^{-1}(g)T(g)$ for
all elements $g$ belonging to the complexification of $G$, $g\in G^{(c)}$, so that $T(g)$ is hermitian where $Z(g)=TrT(g)$. The normalization condition $Tr\rho(g)=1$ is obviously satisfied. In the GCS
case the projection that closes the dynamical equation can be formulated as
follows: We define a tangent vector $v(\rho)$  that satisfies the following
property:

\begin{eqnarray}  \label{tangent}
Tr\{T(a)v(\rho)\}=Tr\{T(a)L\rho\}
\end{eqnarray}
for all $a \in {\cal A}^{(c)}$ and $\rho\in M_{L}$. For most practical applications there
is one and only one tangent vector $v$ for any $\rho$ that satisfies Eq. (\ref
{tangent}). In this case, a well-defined vector field $v(\rho)$ describes the
variational dynamics in the manifold $M_{L}$ of trial density matrices. The physical meaning of Eq. (\ref{tangent}) is clear: If we refer to the operators $%
T(a)$ with $a \in {\cal A}^{(c)}$ as the fundamental operators, the  variational  dynamics
is obtained by the requirement that the dynamical equations hold for the
expectation values of the  fundamental operators. This implies that, similar to the zero temperature case, the variational equation of motion at
finite temperature is derived using the Heisenberg equations of motion.

We assume that the
trial density matrices are represented by finite temperature equilibrium density matrices  with Hamiltonians given by the fundamental operators:
\begin{eqnarray}  \label{trial}
\rho=Z^{-1}\exp[-\beta T(a)]=Z^{-1}T(g)
\end{eqnarray}
for some $a\in {\cal A}^{(c)}$. This implies $g=\exp(\beta a)$ i.e. elements $g\in G$ of the Lie group $G$ can be represented as the exponentials of the corresponding complex Lie algebra elements $a$ and it follows then that $g \in G^{(c)}$, because if ${
\mathcal A}$ is the Lie algebra of the
group $G$, then
${\mathcal A}^{(c)}$ is the Lie algebra that corresponds to $G^{(c)}$.

We conclude this section by demonstrating that this way of closing the
dynamical equation [Eq. (\ref{tangent})] guarantees the compatibility of the
equilibrium and dynamical variational approaches. Using the variational
approach, the equilibrium density matrix can be obtained by finding the
minimum of the Helmholtz free energy

\begin{eqnarray}  \label{therm_pot}
F(\rho)=Tr(H\rho)-\beta^{-1}Tr(\rho\log(\rho))
\end{eqnarray}
among the normalized trial density matrices $\rho \in M_{L}$. The requirement $\delta F=0$
yields:

\begin{eqnarray}  \label{var_eq}
\delta F(\rho)=Tr(H\delta\rho)-\beta^{-1}Tr(\delta\rho \rho)
\end{eqnarray}
for any tangent $\delta\rho$. It follows from Eq. (\ref{tangent}) that $%
\delta\rho=[a,\rho]$ is tangent for any $a \in {\cal A}^{(c)}$. This yields:

\begin{eqnarray}  \label{var_eq_2}
Tr\{T(a)L(\rho_{0})\}&=&Tr\{T(a)[H,\rho_{0}]\}=Tr\{H[T(a),\rho_{0}]\}
\nonumber \\
&=&\beta^{-1}Tr\{\log\rho_{0}[T(a),\rho_{0}]\}
=\beta^{-1}Tr\{T(a)[\log\rho_{0},\rho_{0}]\}=0
\end{eqnarray}
which implies that $v(\rho_{0})=0$. Stated differently, the trial density matrix
which minimizes the free energy is a stationary point of the dynamical
equations.

\section{Variational equations in real space for the contact potential} \label{simple}

For contact interatomic interaction, $\bar{V}({\bf r},{\bf r}^{\prime })
\equiv U_{0}\delta({\bf r}-{\bf r}^{\prime })$, ${U}_0 = \frac{4 \pi
\hbar^{2}a}{m}$, where $a$ is the $s$-wave scattering length and $m$ is the
mass of a single atom. This implies  that the integrations in Eqs. (\ref{z_r}-\ref{kappa_rr}) are removed:

\begin{eqnarray}
i \hbar \frac{d z({\bf r})}{dt} & = & H^{sp}({\bf r}) z({\bf r}) + U_{0} \left
\{ |z({\bf r})|^{2}z({\bf r}) + 2 \rho({\bf r}, {\bf r})z({\bf r}) + \kappa({\bf r}, {\bf %
r})z^{*}({\bf r})  \right \} + V_{f}({\bf r},t) z({\bf r})  \label{positionGPE}\\
i\hbar \frac{d\rho ({\bf r},{\bf r}^{\prime })}{dt} &=& \left [ H^{sp}({\bf r%
}) + 2 U_{0} \tilde{\xi}({\bf r}^{\prime },{\bf r}^{\prime }) \right ] \rho (%
{\bf r},{\bf r}^{\prime }) + U_{0} \tilde{\zeta} ({\bf r}^{\prime },{\bf r}%
^{\prime })\kappa ^{\ast }({\bf r},{\bf r}^{\prime })  \nonumber \\
&-& \left [ H^{sp}({\bf r}^{\prime }) + 2 U_{0} \tilde{\xi}({\bf r},{\bf r})
\right ] \rho ({\bf r},{\bf r}^{\prime }) - U_{0}\tilde{\zeta} ^{\ast }({\bf %
r},{\bf r})\kappa ({\bf r},{\bf r}^{\prime })  \nonumber \\
&+& V_{f}({\bf r},t) {\rho}({\bf r},{\bf r}^{\prime }) - V_{f}({\bf r}',t) {\rho}({\bf r},{\bf r}^{\prime })  \\
i\hbar \frac{d\kappa ({\bf r},{\bf r}^{\prime })}{dt} &=&\left[ H^{sp}({\bf r%
})+2U_{0}\tilde{\xi}({\bf r}^{\prime },{\bf r}^{\prime })\right] \kappa (%
{\bf r},{\bf r}^{\prime })+U_{0}\tilde{\zeta}({\bf r}^{\prime },{\bf r}%
^{\prime })\rho ^{\ast }({\bf r},{\bf r}^{\prime })+U_{0}\tilde{\zeta}({\bf r%
}^{\prime },{\bf r}^{\prime })  \nonumber \\
&+&\left[ H^{sp}({\bf r}^{\prime })+2U_{0}\tilde{\xi}^{*}({\bf r},{\bf r})\right]
\kappa ({\bf r},{\bf r}^{\prime })+U_{0}\tilde{\zeta}({\bf r},{\bf r})\rho (%
{\bf r},{\bf r}^{\prime }) \nonumber \\
&+& V_{f}({\bf r},t) {\kappa}({\bf r},{\bf r}^{\prime }) + V_{f}({\bf r}',t) {\kappa}({\bf r},{\bf r}^{\prime }) ,
\end{eqnarray}
where $\tilde{\xi}({\bf r},{\bf r})$ and $\tilde{\zeta}({\bf r},{\bf r})$ are as given in Eqs. (\ref{xi}-\ref{zeta}).
The equations of motion for $\tilde{\rho}({\bf r},{\bf r}^{\prime}, t)$ and $%
\tilde{\kappa}({\bf r},{\bf r}^{\prime}, t)$ may be written in the compact
form:

\begin{equation}
i\hbar \frac{d{\cal {G}}}{dt}=\Sigma {\cal G}-{\cal G}\Sigma^{\dagger}
\label{positionHFB}
\end{equation}
where we have defined $2 \times 2$ matrices
\begin{equation}
\Sigma ({\bf r,r}^{\prime })=\left(
\begin{array}{cc}
\tilde{h}({\bf r,r}^{\prime }) & \tilde{\Delta}({\bf r^{\prime },r}^{\prime
}) \\
-\tilde{\Delta}^{\ast }({\bf r,r}) & -\tilde{h}^{\ast }({\bf r,r}^{\prime })
\end{array}
\right) \;\;\;\;\;\;{\cal G}({\bf r},{\bf r}^{\prime })=\left(
\begin{array}{cc}
{\rho}({\bf r},{\bf r}^{\prime }) &  {\kappa}({\bf r},{\bf r}%
^{\prime }) \\
{\kappa}^{*}({\bf r},{\bf r}^{\prime }) &  {\rho}^{\ast }({\bf r},%
{\bf r}^{\prime }) + 1
\end{array}
\right) ,
\end{equation}
and
\begin{eqnarray}
\tilde{h}({\bf r,r}^{\prime }) & \equiv &H^{sp}({\bf r}) + V_{f}({\bf r},t) +2U_{0}\tilde{\xi}({\bf r}^{\prime },{\bf r}^{\prime }), \\
\tilde{\Delta}({\bf r,r}) &\equiv & U_{0}\tilde{\zeta}({\bf r},{\bf r}).
\end{eqnarray}
Eqs. (\ref{positionGPE}) and (\ref{positionHFB}) constitute the TDHFB
equations for the contact interatomic potential approximation, in real space\cite{HFB1}.

\section{Zero temperature relationship between $\protect\rho$ and $\protect\kappa$} \label{relrhokappa}

In order to derive the relation between $\rho_{ij}$ and $\kappa_{ij}$ at zero
temperature using the GCS ansatz, we find that it suffices to consider GCS state $| \Omega \rangle$
such that $\langle \Omega | \hat{a}_{i} | \Omega \rangle = 0$ i.e.  Gaussian wave functions centered at $q = 0$. These states form an orbit $M$ of the group $G$ which
corresponds to the algebra generated by $%
\hat{T}^{(z)}_{m n}$ and $\hat{T}^{(\pm)}_{m n}$. We shall introduce a set of functions $%
S^{(z)}_{m n}$, $S^{(\pm)}_{m n}$ on $M$
\begin{equation}
S^{(z)}_{m n} (\Omega) \equiv \langle \Omega | \hat{T}^{(z)}_{m n} | \Omega
\rangle, \;\;\;\;\; S^{(\pm)}_{m n} (\Omega) \equiv \langle \Omega |
\hat{T}^{(\pm)}_{m n} | \Omega \rangle
\end{equation}
and define two sets of auxiliary functions
\begin{eqnarray}
F_{mn} (\Omega) & = & \sum_{\alpha} \left [ S^{(+)}_{m \alpha} S^{(-)}_{\alpha n}
- S^{(z)}_{m \alpha}S^{(z)}_{\alpha n} \right ],   \label{F} \\
G_{mn} (\Omega) & = & \sum_{\alpha} \left [ S^{(z)}_{m \alpha} S^{(+)}_{\alpha n}
- S^{(z)}_{n \alpha} S^{(+)}_{\alpha m} \right ].  \label{G}
\end{eqnarray}

Our aim is to show that  $F_{mn}(\Omega)$ is a constant i.e. its derivatives are zero.  In particular, showing that $F_{mn}(\Omega)  = \delta_{mn}$ and identifying the expectation values $S^{(\pm)}_{m n}$ and $S^{(z)}_{m n}$ in terms of $\rho_{ij}$ and $\kappa_{ij}$ for our example completes the required proof. It is found that the derivatives of $F_{mn}(\Omega)$ are linear combinations of $G_{mn}(\Omega)$, and  therefore it suffices to prove that the auxiliary function $G_{mn}(\Omega)$ is zero for all $m$ and $n$.

We note that since $[\hat{T}^{(+)}_{ij}, \hat{T}^{(+)}_{ke} ] = 0$, the operators $\hat{T}_{ij}^{(+)}$ which are considered as vector fields on $M$ determine a complex structure on $M$.
A function $f$ is
said to be holomorphic if it satisfies the condition $\hat{T}^{(+)}_{m n} f =0$.
Operator $\hat{T}^{(+)}_{m n}$ then represent derivatives in the antiholomorphic
direction.

In particular, the functions $S^{(+)}_{mn}$ constitute a set of
holomorphic coordinates in the vicinity of $\Omega_{0}$ where $\Omega_{0}$
represent states with Gaussian wave functions. It can be
shown that $S^{(+)}_{mn}(\Omega_0) = 0$ while $S^{(z)}_{mn}(\Omega_0) = \delta_{mn}$ so that
\begin{equation}
F_{mn} (\Omega_{0}) = \delta_{mn} . \label{FO}
\end{equation}

A direct calculation yields $\hat{T}^{(+)}_{ij}G_{mn} = 0$ which implies that $G_{mn}$ is holomorphic and can therefore be written as a series in $S^{(+)}_{ij}$ in the vicinity of the point $S^{(+)}_{ij} = 0$ (i.e. $\Omega_{0}$). 
Since $S^{(z)}_{ma}(\Omega) = \delta_{ma}$, it follows from Eq. (\ref{G})  that the expansion of $G_{mn}$ starts with the second-order terms:
\begin{equation}
G_{mn} = \sum_{j=2}^{\infty}  G_{mn}^{(j)} . \label{Gexp}
\end{equation}

We next define the degree of a function $f$ by $\hat{D}f = \hat{{\rm deg}} f \equiv \frac{1}{2} \sum_{j}\hat{T}^{(z)}_{jj} f$. It is clear that $\hat{{\rm deg}} S^{(\pm)}_{ij} = \pm 1$, $\hat{{\rm deg}} S^{(z)}_{ij} = 0$, and $\hat{{\rm deg}} (fg) = =\hat
{{\rm deg}} f + \hat{{\rm deg}}g$. It follows from Eq. (\ref{G}) that $\hat{{\rm deg}} G_{mn} = 1$.  On the other hand Eq. (\ref{Gexp}) implies that $\hat{{\rm deg}}G_{mn}^{(j)} = j$ and therefore contains the degrees of 2 and higher. This implies that
$G_{mn}\equiv 0$ for all $m$ and $n$.

It can be verified that $\hat{T}^{(+)}_{m n} F_{mn}$ is a linear
combination of $G_{ab}$ and hence $\hat{T}^{(+)}_{m n} F_{mn} = 0$.
Similarly, by
conjugating the relation $\hat{T}^{(+)}_{m n} F_{mn} = 0$, $\hat{T}^{(-)}_{m n} F_{mn} =
0$,  implying that $F_{mn} (\Omega)$ is a constant. This, together with
Eqs. (\ref{F}) and (\ref{FO}) imply
\begin{equation}
F_{mn}(\Omega)  = \sum_{\alpha} \left [ S^{(+)}_{m \alpha} S^{(-)}_{\alpha n} - S^{(z)}_{m
\alpha} S^{(z)}_{\alpha n} \right ] = \delta_{mn} .  \label{S1}
\end{equation}
Since $S^{(-)}_{mn} = \kappa_{mn}$ and $S^{(z)}_{mn} = \rho_{mn} + \frac{1}{2%
} \delta_{mn}$ Eq. (\ref{S1}) gives
\begin{equation}
(\rho_{ij} + \frac{1}{2} \delta_{ij} )^{2} - \sum_{p} \kappa_{ip}^{*}
\kappa_{pj} = \frac{1}{4} \delta_{ij} .
\end{equation}
Solving for $\rho_{ij}$ finally yields:
\begin{equation}
\rho_{ij} = \sqrt{\frac{1}{4} \delta_{ij} + \sum_{p} \kappa_{ip}^{*}
\kappa_{pj} } - \frac{1}{2} \delta_{ij}
\end{equation}
or in matrix form,
\begin{equation}
{\bf \rho} = \sqrt{\frac{1}{4} I + {\bf \kappa}^{\dagger} {\bf \kappa} } -
\frac{1}{2} I ,  \label{rhokappa2}
\end{equation}
where $I$ is the unit operator.

Eq. (\ref{rhokappa2}) may also be expanded in a Taylor series as:
\begin{equation}
{\bf \rho} = {\bf \kappa}^{\dagger} {\bf \kappa} - \left ( {\bf \kappa}%
^{\dagger} {\bf \kappa} \right )^{2} + 4 \left ( {\bf \kappa}^{\dagger} {\bf %
\kappa} \right )^{3} - \cdots
\end{equation}

It is possible to gain further insight into the nature of  GCS from the fact that Eq. (\ref{rhokappa2}) holds if the quantum state of the system is a quasiparticle
vacuum state $|0\rangle _{qp}$ such that $\hat{\beta}_{i}(t)|0\rangle
_{qp}=0$ where $\hat{\beta}_{i}(t)$ is the quasiparticle annihilation
operator of the Bogoliubov transformation, $\hat{a}_{i}(t)=\sum_{j\neq
0}U_{ji}\hat{\beta}_{j}(t)+V_{ji}^{\ast }\hat{\beta}_{j}^{\dagger }(t)$. For
a quasiparticle vacuum state, $\rho _{ij}$ and $\kappa _{ij}$ may be
written in terms of matrices $U$ and $V$ as follows:
\begin{equation}
\rho _{ij}=\sum_{p\neq 0}V_{pi}^{\ast }V_{pj}\;\;\;\;{\rm and}\;\;\;\;\kappa
_{ij}=\sum_{p\neq 0}U_{pj}V_{pi}^{\ast }
\end{equation}
which implies
\begin{equation}
\rho ^{2}+\rho =\kappa ^{\dagger }\kappa \;\;\;\;{\rm and}\;\;\;\;\rho
\kappa =\kappa \rho ^{\ast },
\end{equation}
using the orthogonality and symmetry conditions between the matrices $U$ and
$V$, $UU^{\dagger }-VV^{\dagger }=1$ and $UV^{T}-VU^{T}=0$. These relations
can be shown to be identical to Eq. (\ref{rhokappa2}) by solving the
quadratic equation in $\rho $. The quasiparticle vacuum state may therefore be
considered a squeezed state of condensate and non-condensate atoms.

\section{Variational equations in the trap basis} \label{trapTDHFB}

\subsection{TDHFB Equations}

The TDHFB equations in trap basis is given as follows:
\begin{eqnarray}
i\hbar \frac{dz_{i}}{dt} &=&\sum_{j}H_{ij}z_{j}+\sum_{jkl}V_{ijkl}\left[
z_{j}^{\ast }z_{k}z_{l}+2\rho _{jk}z_{l}\right] +\sum_{kl}V_{ijkl}\kappa
_{kl}z_{j}^{\ast } + \sum_{j}E_{ij}z_{j} \label{zdot} \\
i\hbar \frac{d\rho _{ij}}{dt} &=&\sum_{r}\left[ H_{ir}+2\sum_{kl}V_{iklr}%
\left( z_{k}^{\ast }z_{l}+\rho _{lk} \right)  + E_{ir} \right] \rho _{rj}-\sum_{r}\left[
H_{rj}+2\sum_{kl}V_{rjkl}\left( z^{*}_{k}z_{l} + \rho _{lk} \right)  + E_{rj}
\right] \rho _{ir}  \nonumber \\
&&+\sum_{r}\left[ \sum_{kl}V_{irkl}\left( z_{k}z_{l}+\kappa _{kl}\right) %
\right] \kappa _{rj}^{\ast }-\sum_{r}\left[ \sum_{kl}V_{rjkl}\left(
z^{*}_{k}z^{*}_{l}+\kappa^{*}_{kl}\right) \right] \kappa _{ir} \label{rhodot} \\
i\hbar \frac{d\kappa _{ij}}{dt} &=&\sum_{r}\left[ H_{ir}+2\sum_{kl}V_{iklr}%
\left( z_{k}^{\ast }z_{l}+\rho _{lk}\right)  + E_{ir} \right] \kappa _{rj}+\sum_{r}%
\left[ H_{rj}+2\sum_{kl}V_{rjkl}\left( z_{k}z_{l}^{\ast }+\rho
_{lk}^{\ast }\right)  + E_{rj} \right] \kappa _{ir}  \nonumber \\
&&+\sum_{r}\left[ \sum_{kl}V_{irkl}\left( z_{k}z_{l}+\kappa _{kl}\right) %
\right] \rho _{rj}^{\ast }+\sum_{r}\left[ \sum_{kl}V_{rjkl}\left(
z_{k}z_{l}+\kappa _{kl}\right) \right] \rho _{ir}+\sum_{kl}V_{ijkl}\left(
z_{k}z_{l}+\kappa _{kl}\right) .  \label{kappadot}
\end{eqnarray}

The TDHFB equations in real space derived in Section III, Eqs. (\ref{z_r}-\ref{kappa_rr}),  may be transformed to the corresponding trap basis, Eqs. (\ref{zdot}-\ref{kappadot}) in a straightforward manner, using the following relations between the real space basis and the trap basis variables:
\begin{equation}
z({\bf r}) = \sum_{i} z_{i} \phi_{i}({\bf r}),  \;\;\;\; \rho({\bf r}, {\bf r}') = \sum_{ij} \rho_{ij} \phi^{*}_{i}({\bf r})\phi_{j}({\bf r}'), \;\;\;\; \kappa({\bf r}, {\bf r}') = \sum_{ij} \kappa_{ij} \phi_{i}({\bf r})\phi_{j}({\bf r}'),
\end{equation}
along with the definition of the tetradic matrix $V_{ijkl}$
\begin{equation}
V_{ijkl} = \frac{1}{2} \Big [ \mbox{$\langle i
j |$}V\mbox{$| k l \rangle$} + \mbox{$\langle j i |$}V\mbox{$| k l \rangle$} %
\Big ],
\end{equation}
where
\begin{equation}
\langle i j |V | k l \rangle  = \int \! d^{3}%
{\bf r} \, d^{3}{\bf r^{\prime}} \, \phi^{*}_{i}({\bf r})\phi^{*}_{j}({\bf %
r^{\prime}})V({\bf r}-{\bf r^{\prime}})\phi_{k}({\bf r^{\prime}})\phi_{l}(%
{\bf r}) ,  \label{Vijkl}
\end{equation}
with $V({\bf r}-{\bf r^{\prime}})$ being a general interatomic potential.
Under the contact interaction approximation, $V_{ijkl}$ takes a simpler form:
\begin{equation}
V_{ijkl} = \frac{4 \pi \hbar^{2}a}{m}  \int d{\bf r} \phi_{i}^{*}({\bf r})\phi_{j}^{*}({\bf r})\phi_{k}({\bf r})\phi_{l}({\bf r}).
\end{equation}

\subsection{TDHFB at zero temperature}

The TDHFB equations, Eqs. (\ref{zdot}) - (\ref{kappadot}) hold for all
temperatures. However, we have noted that the variables $\rho$ and $\kappa$ are not independent variables for the generalized coherent state ansatz at $T=0$.
$\rho_{ij}$ can therefore be eliminated using the following relation
\begin{equation}
{\bf \rho} = \left [ \sqrt{\frac{1}{4} I + {\bf \kappa}^{\dagger} {\bf \kappa%
} } - \frac{1}{2} I \right ] + \delta {\bf \rho},
\end{equation}
where the function $\delta\rho = 0$ for $T = 0$, the TDHFB equations take
the form:

\begin{eqnarray}
i \hbar \frac{d z}{dt} & = & {\cal H}_{z} z + {\cal H}_{z*} z^{*} + Ez \label{zdotT} \\
i \hbar \frac{d \kappa}{dt} & = & (h \kappa + \kappa h^{*}) + (\left [ \sqrt{%
\frac{1}{4} I + {\bf \kappa}^{\dagger} {\bf \kappa} } - \frac{1}{2} I +
\delta\rho \right ] \Delta + \Delta \left [ \sqrt{\frac{1}{4} I + {\bf \kappa%
}^{\dagger} {\bf \kappa} } - \frac{1}{2} I + \delta\rho \right ] + \Delta ,
\label{kappadotT} \\
i \hbar \frac{d \delta \rho}{dt} & = & [h, \sqrt{\frac{1}{4} I + {\bf \kappa}%
^{\dagger} {\bf \kappa} } - \frac{1}{2} I + \delta\rho] - (\kappa \Delta^{*}
- \Delta \kappa^{*}) - i \hbar \frac{1}{\sqrt{ I + 4{\bf \kappa%
}^{\dagger} {\bf \kappa} } } \left ( \frac{d {\bf \kappa}^{\dagger}}{dt}{\bf %
\kappa} + {\bf \kappa}^{\dagger} \frac{d {\bf \kappa}}{dt} \right )
\label{rhodotT}
\end{eqnarray}
where
\begin{eqnarray}
\left [ {\cal H}_{z} \right ]_{ij} & = & H_{ij} + \sum_{kl} V_{iklj} \left [z^{*}_{k}z_{l} + 2\sqrt{\frac{1}{4} \delta_{lk} + \sum_{m}\kappa^{*}_{lm}
\kappa_{mk} } - \delta_{lk} + 2 \delta\rho_{lk} \right ] \\
\left [ {\cal H}_{z*} \right ]_{ij} & = & \sum_{kl} V_{iklj} \kappa_{kl}  \\
h_{ij} & = & H_{ij} + 2 \sum_{kl} V_{iklj}
\left [ z_{k}^{*}z_{l} + \sqrt{\frac{1}{4} \delta_{lk} +
\sum_{m}\kappa^{*}_{lm} \kappa_{mk} } - \frac{1}{2} \delta_{lk} +
\delta\rho_{lk} \right ] + E_{ij} \\
\Delta_{ij} & = & \sum_{kl} V_{ijkl} \left[ z_{k}
z_{l} + \kappa_{kl} \right] . \\
\end{eqnarray}
At $T = 0$, $\delta \rho = 0$ and $\rho = \sqrt{\frac{1}{4} I +
\kappa^{\dagger} \kappa } - \frac{1}{2} I $. By direct differentiation of  $\rho$
\begin{equation}
i \hbar \frac{d \rho }{dt}  =  i \hbar \frac{1}{\sqrt{I + 4{\bf \kappa}%
^{\dagger} {\bf \kappa} }} \left ( \frac{d {\bf \kappa}^{\dagger}}{dt}{\bf %
\kappa} + {\bf \kappa}^{\dagger} \frac{d {\bf \kappa}}{dt} \right ) ,
\label{D6}
\end{equation}
while  Eq. (\ref{rhodot}) implies
\begin{equation}
i \hbar \frac{d \rho }{dt}  =   [h, \sqrt{\frac{1}{4} I + {\bf \kappa}^{\dagger} {\bf \kappa} } -
\frac{1}{2} I ] - (\kappa \Delta^{*} - \Delta \kappa^{*}).  \label{D7}
\end{equation}
Since Eqs.(\ref{D6}) and (\ref{D7}) are equivalent, both the left and the right hand side of Eq. (\ref{rhodotT}) are zero at $T=0$, $\delta \rho = 0$ i.e. the only independent equations to be solved are Eqs. (\ref{zdotT}) and (\ref{kappadotT}).

For comparison, we note that the GPE, which is a zero temperature theory for a coherent state ansatz, is simply obtained from Eq. (\ref{zdotT}) by setting $\rho_{ij}  = \kappa_{ij} = 0$:
\begin{equation}
i \hbar \frac{d z_{i}}{dt} = \sum_{j} \left [ H_{ij} + \sum_{kl} V_{iklj} z^{*}_{k}z_{l}  + E_{ij} \right ] z_{j}.
\end{equation}

\end{document}